\documentclass[aps,prl,twocolumn,groupedaddress]{revtex4}

\usepackage[dvips]{graphicx}
\usepackage[dvips]{color}

\begin{document}
\definecolor{Red}{rgb}{1,0,0}

\title{Mapping markets to the statistical mechanics: the derivatives act against the self-regulation of stock market
}
\author{David B. Saakian$^{1,2,3}$}

\affiliation{$^1$Yerevan Physics Institute, Alikhanian Brothers St.
2, Yerevan 375036, Armenia}
\affiliation{$^2$Institute of Physics,
Academia Sinica, Nankang, Taipei 11529, Taiwan}

\affiliation{$^3$National Center for Theoretical Sciences:Physics
Division, National Taiwan University, Taipei 10617, Taiwan}


\date{\today}

\begin{abstract}
Mapping the economy to the some statistical physics models we get
strong indications that, in contrary to the pure stock market, the
stock market with derivatives could not self-regulate.

\end{abstract}
 \maketitle


\begin{center}
\begin{table}[tbhp]
\begin{tabular}{l@{\hspace{4mm}}c@{\hspace{4mm}}c}
  & Economics.  &   Statistical physics.  \\
 & Money. &   Energy.  \\
& Value. &   Free energy.  \\
& The capitalization of stocks. &   Work. \\
 & Volatility. &   Local temperature.   \\
& Market's Kolmogorov &  Entropy \\
& complexity.&  \\
& Infinitely effective market. &   Hamiltonian \\
&  &  Statistical mechanics. \\
& Self-regulating real market.  &  Thermostat, 0-th law,2-nd  \\
&  &law of thermodynamics. \\
&Accidental arbitrage. & Maxwell Demon. \\
& Strong usage of derivatives, &  Strong Born-Openheimer \\
& crisis. &  interaction \cite{al}, violation of \\
& & thermodynamic laws. \\
\end{tabular}
\caption{ The correspondence between economics and statistical
physics.}
\end{table}
\end{center}
{\bf Introduction}. How can we understand the crisis? Shall we
revise the whole economic theory of highly effective self-regulating
market with "infinitely rational agents" in favor of more modern
theories (behaviorial economics")? Our idea is that there is no need
for a radical change in economic theory.  We should just borrow more
ideas from statistical physics, and the reflection concept of
Marxist philosophy.

 In recent decades it has been well realized that some aspects of financial
 markets could be analyzed well with the methods of statistical
 physics, especially the idea of scaling for the fat tails of distributions \cite{st00}.
 The scaling is identified in statistical physics with the
 universality of critical phenomena near the second order phase
 transition point.
 This rather technical approach was certainly useful.
 Nevertheless, statistical physics is much richer discipline than
 only second order phase transition theory.
 In \cite{fa02} has been observed the similarity of money and free
 energy. V. P. Maslov identified the entropy of market with its
 complexity \cite{ma08}, and I just borrow his idea for the Table 1, giving the
 correspondence between statistical physics and economics.
A fruitful  mapping of the wealth distribution and international
trade to thermodynamic has been done in \cite{mi05,ya09}. The
thermodynamic approach was very successful here, because in this
narrow field there is an equivalent of the first law (the
conservation of money). We are interested in a more general
situation.

 In our article \cite{sa05} we tried to identify different classes
 of universality of complex systems, rather than consider only
 the second order phase transitions.
 One can use several variables to identify the universality class (the  systems from the same class should share the same
 criteria): a. The subdominant behavior of free energy or some entropy; b. intrinsic information
-theoretical aspects (heterogeneity of agents,
ferromagnetic-antiferromagnetic couplings, gauge
 invariance).
 These concepts could be useful for investigation of simple
 markets.

 Let as analyze the economics using the further ideas borrowed from statistical physics and philosophy.
  In Table 1, we give some correspondence between two areas.
The connection  between the II law of thermodynamics and
no-arbitrage property of market is  known. Perhaps we should take
Thomson's formulation: it is impossible get a work (income) from a
system in the equilibrium.

 A fundamental concept for analyzing any serious phenomenon in physics and in interdisciplinary
 science is the concept of reflection, well realized in the Marxist
 philosophy.

 There is an objective reality, identified sometimes with the
 substance of materia, and its reflection- the subjective reality.
 It is not an abstract philosophical concept. On the contrary, it is highly
 concrete and useful for applications in statistical physics.

 Another concept we need, is an amount of motion invented by philosophers De-Cartesius and Leibnitz.
  It has been identified in
 classical mechanics first with momentum, later with
 energy.

{\bf Statistical physics.}
 The reflection and amount of motion are explicit in statistical physics.
 There is a hierarchy of motions:
 at the background level - a microscopic motion of molecules with some
 energy.
 This is an objective reality.
The system quickly goes to some equilibrium and it is possible to
define the temperature (0-th law of thermodynamic).

 At the next level, we have a thermodynamics (with an observer, as has been realized
 by W. Heisenberg) with free energy. Using the idea of Gibbs, one
 calculates the partition (a probability like quantity), then, taking
 its logarithm, the free energy.
 The free energy is the manageable amount of motion on a macroscopic
 level.

{\bf Advanced case.} There are physical systems, where the free
energy of one system is
 a partition of another system
 \cite{ts96}
 .
In spin glasses there is a simple description of a hierarchy when
there is a fast relaxation, and a quite complicated in case of
relaxation with the same rates on both levels of hierarchy.
 There is a hierarchy of time scales, or hierarchy of interactions.
Sometimes there are some problems with ergodicity (spin-glass phase)
but even for such a situation we have valid thermodynamical
expressions for a free energy, therefore the "amount of motion" is
well defined \cite{sh93},\cite{sa99}.

In \cite{al} has been carefully investigated the statistical physics
problem of interaction between slow and fast variables.
 It has been
considered two level hierarchy system. If  A influenced B, and there
is amount of motion (energy) for B, it is possible to define well
the amount of motion for A (free energy), and the thermodynamic
approach with corresponding free energy is valid for A. It is
possible to have an alternative situation, when there is a backward
influence of B to A. Such case arose when the interaction energy
between A and B is strong as the energy of A, and the effective
interaction A-B at some moment of time depends from the state of A
in the previous period. Then, as a result, the system is becoming
non-ergodic and it is impossible to construct proper thermodynamics
(the second Law is broken).

{\bf Self-regulating market. } Now we have goods. They have some
value (amount of motion), this can be expressed via money. The
companies have some values,  expressed in the stocks. Thus we have a
simple case of reflection, expressed as a reflection from an
objective reality (goods, factory) to the subjective reality. The
money is a simple form of motion.
Till now everything is similar to the theory of thermodynamics. The
idea of a self-regulating market is quite reasonable from the
statistical physics point of view. It is similar to the 0-th law of
thermodynamics with the idea of thermostat and Gibbs distribution.

{\bf The heterogeneity of agents.} While this aspect of economics is
usually ignored, there are special situations when the heterogeneity
of agents is crucial, as has been suggested in \cite{lu09}, see an
econophysics example \cite{ch07}. In statistical physics it is well
known that the ensemble of systems could not be replaced by a single
system with "typical" coupling in case of spin-glass phase, and the
Berry phase could not be revealed by an electronic Hamiltonian with
a typical position of atoms. In \cite{sa05} has been assumed to
identify the financial market data with the phase transition point
between ferromagnetic and spin-glass phases, which could not be
described by a homogenous set of agents. This is a transition point
strongly influenced by disorder\cite{do95}. In principle one could
look for different such points from different universality classes
to capture the market data.
 The heterogeneity of agents is
one of three principal points to identify the models: first and most
important is the character of reflection, second: the subdominant
term of free energy, and third: intrinsic information transition
mechanisms (heterogeneity of agents, a gauge symmetry for the
ensemble of quenched couplings and magnetization). If the model is
not chosen according to these three criteria, then it does not
belong to the universality class of market, and  could not be
considered as a realistic one.

{\bf Financial derivatives.} They are subjective realities,
connected with underlying assets (stocks) as objective realities.
The relaxation time is very short, there is no hierarchy. This is a
crucial point. When there is a strong backward interaction (both
hedging and speculation), the situation highly resembles the
statistical physics problem of \cite{al} with strong backward
interaction. Therefore, in case of strong backward interaction
between underlying assets and derivatives the market should not
self-regulate.

 The Maxwell demon phenomenon is well known in statistical physics.
 It is as improbable there, as arbitrage situation in markets.
 In case of extensive organization
of Maxwell demon like situations the whole statistical physics will
cease to work, like the crash of the market in case of permanent
arbitrage.

It is not clear by statistical physics analogy whether the
derivatives create arbitrage situation, but what the analogy tells
is that the unrestricted usage of derivatives could create strong
non-ergodicity.
 Such
derivatives are something different than an amount of motion,
therefore they could not self-regulate. The usefulness of
unrestricted usage of derivatives has been questioned in
\cite{lu09}. The authors of \cite{lu09} realized that new dangerous
connections could be created by derivatives.

{\bf Hot design}.  For the classic problem of gambler's ruin it is
well known that unlimited optimization of total return leads to ruin
with probability one \cite{co91}. Carlson and Doyle
 have proposed a model \cite{ca99} for designed systems which they call
"highly optimized tolerance," or HOT design, to avoid some dangers
(faults) in an optimal way. What was found in \cite{ca99}, the HOT
design systems are robust to perturbations they were designed to
handle, yet are fragile to rare perturbations and design flaws
\cite{fa02a}. The economists claim the usefulness of derivatives to
avoid some fluctuations of markets. It is funny that the derivatives
bring, as a good illustration of  HOT design idea, to the crash of a
whole economy. The idea of economics that the derivatives are used
only for hedging contradicts to the practice of markets (due to
problem with liquidity of market). A solution of the problem is the
restricted optimization only \cite{fa02a}.

{\bf Conclusions.} While statistical physics has been applied to
financial market in many articles, this is the first work with deep
application of statistical physics to financial markets with
derivatives. To model correctly the economy, we should consider the
statistical physics models with the same character of complexity.
 A perfect market economy
resembles the statistical physics with a Hamiltonian type of
interactions, and the self-regulating market is a system with a
thermostat (thermodynamic state). While widely questioned in recent
years \cite{bu08}, the idea of self regulating market for real
economy with partial rationality of agents is not a worse hypothesis
than the usage of statistical physics without proving the ergodicity
or mixing property.

We analyzed the market with derivatives using the analogy with the
model in \cite{al}.
  In \cite{al} a strong
backward interaction between slow and fast variable is considered
which is similar to unrestricted usage of derivatives in markets.
Using the Liouville theorem, in \cite{al} is proved the violation of
thermodynamics second law. While the similarity of market with
statistical physics is rather restricted (there is no an equivalent
of the thermodynamics First law and the Liuoville theorem),
 the result of \cite{al} is a strong
indication  that financial markets with unrestricted usage of
derivatives could not self-regulate. We have the same conclusion
from the analogy with HOT design problem \cite{ca99}: trying
maximally neutralize some risks in complex system (for hedgers in
market), one creates a danger for the global crash of the system. It
is important that HOT design problem exist for a rather general
situation when one tends to unlimited optimization (certainly the
case of stocks market with derivatives). The danger of unrestricted
optimization (sybaritism versus unrestricted usage of derivatives in
our case) has been well recognized long ago before \cite{ca99,fa02a}
by Epicures:
 " No pleasure is a bad
thing in itself, but the things which produce certain pleasures
entail disturbances many times greater than the pleasures
themselves",\cite{ep}. There are no any known example of
self-regulating system with double reflection (a process like to the
transition from energy to free energy) with strong interaction.  Why
markets with derivatives should be an exception? Therefore should be
restriction against their unrestricted usage, like the "constrained
optimization" design to avoid the global crash of the system
\cite{fa02a}.

Even when the systems are entirely Hamiltonian, they do not
necessarily  bring to the thermodynamics state with the Gibbsian
distribution. Furthermore, the crises in economy happened before the
invention of derivatives. Nevertheless, the constraint against the
unrestricted usage of derivatives  removes a quite realistic danger
of crisis. In any case, there is a need to perform very careful
testing of microscopic market models, considering their relaxation
dynamics \cite{ch07} to verify our criticism of the myth of
economics about self-regulating stock market with derivatives.


{\bf Acknowledgment}. I thank  A. Allahverdyan, H. Kechejian, Th.
Lux, Z. Struzik for discussions. The work was supported by
Volkswagenstiftung grant "Quantum Thermodynamics'',
 National Center for Theoretical Sciences
in Taiwan and Academia Sinica (Taiwan),Grant No. AS-95-TP-A07.

\end{document}